\documentclass[12pt]{article}
\usepackage{amsmath,amssymb}

\usepackage{graphicx}
\usepackage{wrapfig}

\usepackage{hyperref}
\usepackage{cite}

\def\beq{\begin{eqnarray}}
\def\eeq{\end{eqnarray}}

\newcommand{\Tr}{\,\mathrm{Tr}\,}            % already defined

\newcommand{\be}{\begin{equation}}
\newcommand{\ee}{\end{equation}}
\newcommand{\bea}{\begin{eqnarray}}
\newcommand{\eea}{\end{eqnarray}}
\newcommand{\bg}{\begin{gather}}

\newcommand{\bseq}{\begin{subequations}}
\newcommand{\eseq}{\end{subequations}}

\renewcommand{\ln}{\mathop{\rm ln}\nolimits}

\def\be{\begin{eqnarray}}
\def\ee{\end{eqnarray}}
\def\lb{\label}

\begin{document}

\title{\textbf{
Metric Redefinition and UV Divergences in Quantum Einstein Gravity}}

\vspace{2cm}
\author{ \textbf{ Sergey N. Solodukhin }} %\copyright

\date{}
\maketitle
\begin{center}
  \hspace{-0mm}
  \emph{  Laboratoire de Math\'ematiques et Physique Th\'eorique  CNRS-UMR
7350 }\\
  \emph{F\'ed\'eration Denis Poisson, Universit\'e Fran\c cois-Rabelais Tours,  }\\
  \emph{Parc de Grandmont, 37200 Tours, France}
\end{center}

{\vspace{-11cm}
\begin{flushright}
%preprint
\end{flushright}
\vspace{11cm}
}

%\hfill{\tt IUB-TH/***}\\\mbox{} \\
%\twocolumn[\hsize\textwidth\columnwidth\hsize\csname
%@twocolumnfalse\endcsname

%\maketitle \thispagestyle{empty}% \vspace*{.5cm}

\begin{abstract}
\noindent { 
I formulate several statements demonstrating that the local metric redefinition can be used to reduce the UV divergences present
in the quantum action for the  Einstein gravity in $d=4$ dimensions. 
In its most general form, the proposal is that any UV divergences in the quantum action  can be removed by an appropriate
field re-definition and a renormalization of cosmological constant.}

%\noindent {PACS: 04.70Dy, 04.60.Kz, 11.25.Hf }}
\end{abstract}
%\vskip 2.pc
%\maketitle

\vskip 2 cm
\noindent
\rule{7.7 cm}{.5 pt}\\
\noindent 
\noindent
\noindent ~~~ {\footnotesize e-mail:  Sergey.Solodukhin@lmpt.univ-tours.fr}

%\newpage
%    \tableofcontents
\pagebreak

\newpage

%\section{ Introduction}
Since the early days of Quantum Field Theory it is known that  all theories can be divided on 
two classes: those with dimensionless coupling constants and the theories in which the coupling is dimensionful.
The theories of the first class have that advantage that they may produce only few possible UV divergent terms so that the UV divergences in these theories may be hidden in the renormalization of a finite number of physical
parameters. The theories, such as QED, in which this procedure works are called renormalizable. The theories of the second class 
are obviously non-renormalizable since a priori there exist an infinite number of possible UV divergent terms.

The Einstein gravity is a theory of the second type. Restricting to maximum two derivatives of metric in the action 
\be
W_E=-\frac{1}{16\pi G}\int_{{\cal{M}}^d}\sqrt{g}(R-2\Lambda) 
\lb{-1}
\ee
one finds that there are at most two dimensionful constants: Newton's constant $G$ and cosmological constant $\Lambda$.
The quantum theory of gravity has a long history which has started with the works of Rosenfeld \cite{Rosenfeld} and Bronstein \cite{Bronstein:2012zz}. 
The modern part of it was developed in the works of Arnowitt, Deser and Misner \cite{Arnowitt:1962hi},  Bryce DeWitt \cite{DeWitt:1967yk} and  't Hooft and Veltman \cite{'tHooft:1974bx}. In  \cite{'tHooft:1974bx} it was  calculated the one-loop UV divergent term. In the dimensional regularization their result is
\be
\Gamma_{(1)}= \frac{1}{(d-4)}\int_{{\cal M}^4} \sqrt{g}(\frac{1}{120}R^2+\frac{7}{20}R_{\mu\nu}R^{\mu\nu})\, .
\lb{0}
\ee
This result indicates that the theory is finite on-shell, $R_{\mu\nu}=0$, provided cosmological constant $\Lambda=0$. This is due to absence of the Riemann tensor term in the one-loop divergence (\ref{0}). 
In fact this is an accident of four dimensions. The term $R_{\alpha\beta\mu\nu}^2$, which is a priori present in the one-loop divergence,  is re-expressed in terms of
$R^2$ and $R^2_{\mu\nu}$ and the Gauss-Bonnet term, the latter after integration produces a topological invariant. In higher dimensions, $d\geq 6$, this mechanism is no more in place and the Riemann tensor shows up already in the one-loop UV divergence, see for instance \cite{vanNieuwenhuizen:1976vb}. This example indicates that the appearance of the Riemann tensor alone, without any contractions to Ricci tensor or its derivatives, is the main  obstruction to the renormalizability of the Quantum Gravity. Indeed, already in two loops such a term has been indeed detected by Goroff and Sagnotti \cite{Goroff:1985th} and later confirmed by van de Ven\cite{vandeVen:1991gw},
\be
\Gamma_{(2)}\sim \frac{G}{(d-4)}\int_{{\cal M}^4} \sqrt{g}R_{\alpha\beta}^{\ \ \mu\nu}R_{\mu\nu}^{\ \ \sigma\rho}R_{\sigma\rho}^{\ \ \alpha\beta}\, .
\lb{00}
\ee
(The double poles in two loops and their vanishing on-shell  were analyzed in \cite{BV}.)
The appearance of similar terms in higher loops is not a priori forbidden by any symmetry so that the issue of the Riemann tensor is indeed the key
point in the non-renormalizability of the Einstein gravity. In the presence of non-vanishing cosmological constant $\Lambda$ the argumentation stays the same, see \cite{Gibbons:1978ji}, \cite{Christensen:1979iy}, \cite{Fradkin:1982kf}.
The only difference is that the on-shell condition $R_{\mu\nu}=\Lambda g_{\mu\nu}$ does not imply that the one-loop UV divergent term is nil. 
However, the divergence which remains can be absorbed in the renormalization of the cosmological constant $\Lambda$. The presence of the two-loop term (\ref{00}), however, still prevents the theory from being
renormalizable.

The recent progress in computing the higher loops in supergravity did not actually change much  this story, as far as   the Einstein gravity is concerned.
Although,  there have been found some unexpected cancellations in   the higher loop diagrams \cite{Bern:2007xj}.

The main idea pursued in this paper is to use a field redefinition of the general form
\be
g_{\mu\nu}(x,\epsilon)=g_{\mu\nu}(x)+\sum_{k}\alpha_k(\epsilon ) g^{(k)}_{\mu\nu}(x)\, ,
\lb{1}
\ee
where $\alpha_k(\epsilon)$ are some functions of $\epsilon$, a UV cut-off, and try to choose functions $g^{(k)}_{\mu\nu}(x)$ properly to remove the UV divergent terms in the Quantum Gravity action. The physical meaning is attached to $g_{\mu\nu}(x)$. 

%Using the standard terminology, $g_{\mu\nu}(x,\epsilon)$ is a bare metric while $g_{\mu\nu}(x)$ is the renormalized metric.

\bigskip

There have been some inspirations for the present work. 

\medskip 

\noindent {\it Earlier work of D. Kazakov.} The first and  main inspiration is the old unpublished  work of D. Kazakov \cite{Kazakov:1987ej}
in which he has proposed to use a field redefinition of the type (\ref{1}) and then, by an appropriate choice of $g^{(k)}_{\mu\nu}(x)$, remove all UV divergences in the Quantum Gravity action.
He considered the dimensional regularization so that in this case $\epsilon=(d-4)$ is dimensionless.  The concrete mechanism consists in mutual cancellation 
between $1/\epsilon^{n}$ divergent terms, by the renormalization group related to the one-loop  $1/\epsilon$ divergence, and the higher loop terms $G^{k}/\epsilon^{n+2k}$.
The cancellation condition boils down to certain differential equations on functions $g^{(k)}_{\mu\nu}(x)$. At least in principle, the appropriate functions  $g^{(k)}_{\mu\nu}(x)$ can be found although they occur to be non-local functions of the metric $g_{\mu\nu}(x)$. In the approach developed below the corresponding equations are algebraic so that the terms in the expansion (\ref{1}) are local functions of the metric $g_{\mu\nu}(x)$.

\medskip

\noindent{\it Similarity to  geometric Ricci flow.} The other inspiration is geometrical. In many aspects the Ricci flow
\be
\partial_\lambda g_{\mu\nu}(x,\lambda)=-R_{\mu\nu}
\lb{2}
\ee
is analogous to the renormalization group equation. Curiously, we find that under this flow the volume and the Einstein-Hilbert term change as follows
\be
\partial_\lambda \int_{{\cal M}^d} \sqrt{g}=-\frac{1}{2}\int_{{\cal M}^d} \sqrt{g}R\, , \ \   \partial_\lambda\int_{{\cal M}^d} \sqrt{g}R=\int_{{\cal M}^d}\sqrt{g}(R_{\mu\nu}R^{\mu\nu}-\frac{1}{2}R^2)\, .
\lb{3}
\ee
The second equation in (\ref{3})  is suspiciously similar to the one-loop UV divergent term (\ref{0}).  The difference in the relative factors  can be cured  
by adding to the Ricci flow (\ref{2}) a term proportional to $g_{\mu\nu}R$ with appropriate factor.  For small $\lambda$, equation (\ref{2}) can be solved as $g_{\mu\nu}(x,\lambda)=g^{(0)}_{\mu\nu}(x)-\lambda R_{\mu\nu}+..$. Identifying $\lambda$ with the appropriate function of the UV cut-off $\epsilon$ we arrive at a field redefinition of the type (\ref{1}). On the other hand, the both equations in (\ref{3}) show that the higher curvature terms can be obtained by differentiating the appropriate number of times the volume with respect to
parameter $\lambda$. This observation illustrates our main point in the paper that the UV divergences, even infinite number of them,  can be  ``hidden'' into the volume term.

\medskip

\noindent {\it Work of E. Witten on $3d$ gravity \cite{Witten:2007kt}.} In $d=3$ dimensions the Riemann tensor is expressed in terms of the Ricci tensor and Ricci scalar
so that the issue of the Riemann tensor in the UV divergent terms does not arise. Although the infinite number of potential counter terms still exists, and the theory appears to be non-renormalizable, 
all of them are constructed in terms of the Ricci tensor and its derivatives. So that these terms can be removed by a field redefinition of the type (\ref{1}), $g_{\mu\nu}\rightarrow g_{\mu\nu}+aR_{\mu\nu}+..~$.
What remains then is to simply  renormalize the cosmological constant. Therefore, as is pointed out in \cite{Witten:2007kt}, ``any divergences in perturbation
theory can be removed by a field redefinition and a renormalization of $l^2$'' ($1/l^2$ is the cosmological constant).  What we want to show in this paper is that exactly this statement
is true in $d=4$ (and higher) dimensions. For that we have to address properly the issue of the Riemann tensor. 
We shall not assume any field equations to be satisfied so that our approach is off-shell (the on-shell condition, however, can be always imposed as we comment later in the note).

Interestingly, the resolution of the problem of the Riemann tensor  can not be done if we do not take into account  the cosmological constant. 
In order to illustrate our point let us start with a particular form (a more general form will be considered below) of   the UV divergences neglecting, in particular, the divergences in the cosmological constant.
In $d=4$ dimensions we have
\be
\Gamma_{div}(\epsilon)=\frac{a_1}{\epsilon^{2}}\int_{{\cal{M}}^4}\sqrt{g}R+\ln\epsilon \sum_{k\ge 0} G^k \int_{{\cal{M}}^4}\sqrt{g} Z^{(k)}(x)\, ,
\lb{4}
\ee
where $Z_{(k)}$ are polynomials of degree $k+2$,  each power of the Riemann tensor or its contraction is counted as degree 1   while the degree is  $1/2$ for each covariant derivative of the Riemann tensor.
It is important for our construction that we are using a UV regularization with dimensionful cut-off $\epsilon$ and include the power-law divergences as well as logarithmic. For the logarithmic term the relation to dimensional regularization is as follows: $\ln\epsilon\sim \frac{1}{d-4}$.

Now, let us re-define the metric $g_{\mu\nu}(\epsilon, x)$ as follows
\be
g_{\mu\nu}(\epsilon, x)=g_{\mu\nu}(x)+\epsilon^2\ln\epsilon \sum_{k\geq 0}h^{(k)}_{\mu\nu}(x)\, .
\lb{5}
\ee
We then formulate our first two statements.

\bigskip

\noindent {\it Statement A:} For any divergences produced by terms $Z^{(k)}$ in (\ref{4}) that contain at least one power of the Ricci tensor, $Z_{(k)}=R_{\mu\nu}Y^{\mu\nu}_{(k)}$, one can 
find $h^{(k)}_{\mu\nu}(x)$  such that, after substitution of (\ref{5}) in (\ref{4}) the corresponding UV divergent terms  cancel. The condition for the cancellation is
\be
a_{1}E_{\mu\nu}h^{\mu\nu}_{(k)}=G^k~R_{\mu\nu}Y^{\mu\nu}_{(k)}\, , \ \  E_{\mu\nu}=R_{\mu\nu}-\frac{1}{2}g_{\mu\nu} R\, .
\lb{6}
\ee
This can be solved as follows\footnote{This solution is up to a tensor $\psi_{\mu\nu}(x)$  orthogonal to the Einstein tensor, $\Tr(E\psi)=0$.
Although such a tensor $\psi_{\mu\nu}$ may exist we did not manage to find an example in the class of local tensors constructed from the curvature. 
}
\be
h_{\mu\nu}^{(k)}=\frac{G^k}{a_1}(Y_{\mu\nu}^{(k)}-\frac{1}{2}g_{\mu\nu} \Tr Y^{(k)})\, ,
\lb{7}
\ee
where the trace is computed with respect to the physical metric $g_{\mu\nu}$. 

\medskip

Clearly, if $Z^{(k)}$ contains the Riemann tensor only and the metric re-definition is in the class of analytic functions  then this mechanism does not work. However, in a class of more general, non-polynomial, functions of curvature the appropriate $h^{(k)}_{\mu\nu}$ can be found in more then one way.

\bigskip

\noindent {\it Statement B.} Any divergences produced by terms $Z_{(k)}$ which contain only the Riemann tensor and its covariant derivatives  can be removed by the field redefinition
(\ref{5}) with $h_{\mu\nu}^{(k)}$ taking one of the following forms:
\be
h^{(k)}_{\mu\nu}=- \frac{1}{a_1R} \, g_{\mu\nu} G^k Z^{(k)}
\lb{8}
\ee
or
\be
h^{(k)}_{\mu\nu}=\frac{1}{a_1X}(\alpha R_{\mu\nu}+\beta g_{\mu\nu}R) G^kZ^{(k)}\,  , \ \ X=\alpha R_{\mu\nu}R^{\mu\nu}+(2\beta-\frac{\alpha}{2}) R^2\, .
\lb{9}
\ee
This, however, may not be fully satisfactory since the re-definition (\ref{5}) with (\ref{8}) or (\ref{9}) is not well-defined near the Ricci flat metrics.

The other point is that we so far ignored the UV divergence of cosmological constant. 
Indeed, in general, the UV divergent action includes a  term  without derivatives of the metric,
\be
\Gamma'_{div}(\epsilon)=\frac{a_0}{\epsilon^4} \int_{{\cal{M}}^4} \sqrt{g}+\frac{a_1}{\epsilon^{2}}\int_{{\cal{M}}^4}\sqrt{g}R+\ln\epsilon \sum_{k\ge 0} G^k \int_{{\cal{M}}^4}\sqrt{g} Z^{(k)}(x)\, .
\lb{4-1}
\ee
At first sight the presence of the cosmological constant term spoils everything since under the redefinition (\ref{5}) it produces a new divergent term proportional to $\epsilon^{-2} \ln\epsilon$ which can not be removed by a  modification of (\ref{5}). However, the presence of the cosmological constant offers a new, much more  interesting, possibility to cancel any UV divergences, including those that depend on the Riemann tensor only. This can be seen from the following Statement.

\bigskip

\noindent {\it Statement C.} Any UV divergences, accept the leading one, in (\ref{4-1}) can be removed by field redefinition  
\be
g_{\mu\nu}(\epsilon, x)=g_{\mu\nu}(x)+ \frac{2a_1}{a_0}\epsilon^2 f_{\mu\nu}+2a_0^{-1}\epsilon^4\ln\epsilon \sum_{k\ge 0}G^kh^{(k)}_{\mu\nu}(x)\, ,
\lb{5-1}
\ee
with the only conditions that
\be
\Tr f=-R \, , \ \ \ \ \ \Tr h^{(k)}=-Z^{(k)}\, , \ k\geq 0\, .
\lb{5-2}
\ee
It should be noted that the solution of (\ref{5-2}), 
\be
f_{\mu\nu}=-\frac{1}{4}g_{\mu\nu}(x)R+\phi_{\mu\nu}(x)\, , \ \ h^{(k)}_{\mu\nu}=-\frac{1}{4}g_{\mu\nu}(x)Z^{(k)}+\phi^{(k)}_{\mu\nu}(x)\, ,
\lb{5-3}
\ee
is not unique, it is determined up to a traceless tensor, $\phi_{\mu\nu}\sim R_{\mu\nu}-\frac{1}{4}g_{\mu\nu}R$ and  $\phi^{(k)}_{\mu\nu}(x)$, $k\geq 0$. In the class of tensors, local covariant functions of metric, at each order $k+2$ there exists a finite number of possible structures for $\phi^{(k)}_{\mu\nu}$. 

\bigskip

It is important to note that this statement is valid for {\it any} invariants $Z^{(k)}$ including those 
which depend on the Riemann tensor only. In particular, the 2-loop divergent term (\ref{00}) can be removed by the field redefinition (\ref{5-1})-(\ref{5-2}).
It should be noted that the redefinition (\ref{5-1}) does not produce any new divergences since any variation of the curvature dependent terms in the action
produces either vanishing or finite in the limit of small $\epsilon$ result. 
After the field redefinition the only UV divergence which remains is the leading one, $1/\epsilon^4$. It can be removed by 
the subsequent renormalization of the cosmological constant.  Thus, our conclusion is that in $d=4$ any UV divergences in the Quantum Einstein Gravity 
can be removed by a field redefinition and a renormalization of cosmological constant. 

The above statement may be generalized for a more general form of the UV divergences. Indeed, we can not exclude,  in the presence of a dimensionful cut-off $\epsilon$,
the appearance of extra contributions due to powers of parameter $z=G\epsilon^{-2}$ to each term in (\ref{4-1}). Thus, the most general form for the UV divergent part of the action,
\be
\Gamma''_{div}(\epsilon)=\frac{a_0(z)}{\epsilon^4} \int_{{\cal{M}}^4} \sqrt{g}+\frac{a_1(z)}{\epsilon^{2}}\int_{{\cal{M}}^4}\sqrt{g}R+\sum_{k,l\ge 0}(\ln\epsilon )^{l+1} \mu_{k,l}(z)G^k \int_{{\cal{M}}^4}\sqrt{g} Z^{(k,l)}(x)\, ,
\lb{6-1}
\ee
includes some functions $a_0(z)$, $a_1(z)$ and $\mu_{k,l}(z)$ of variable $z=G\epsilon^{-2}$  as well as the higher logarithmic terms,  $Z^{(k,l)}$ are local functions of curvature and its covariant derivatives. Our assumption is that the UV divergence of cosmological constant is still the dominant one so that
\be
\epsilon^2 a_1(z)a_0(z)^{-1}\rightarrow 0\, , \ \ \epsilon^4(\ln\epsilon)^{l+1} ~\mu_{k,l}(z)a_0^{-1}(z)\rightarrow 0\ , \ \  {\rm if}\  \epsilon \rightarrow 0\, .
\lb{7'}
\ee
Since we are always free to change our UV cut-off, the re-parametrization $\epsilon\rightarrow f(\epsilon)$ can be used to impose $a_0(\epsilon)=1$ (assuming $a_0(\epsilon)>0$).
Without loss of generality we shall assume this value for the function $a_0$. 
Assuming, for  purposes of illustration,  a power law at small $\epsilon$, the conditions (\ref{7'})  imply that
\be
a_1(\epsilon)\sim \epsilon^{\lambda-2}\, , \ \ \ \mu_{k,l}(\epsilon)\sim \epsilon^{\gamma_{k,l}-4}\, , \ \ \  \lambda>0\, , \ \ \gamma_{k,l}>0\, .
\lb{7''}
\ee

In this  case we have a more general statement.

\bigskip

\noindent {\it Statement D.} Any UV divergences, accept the leading one, in (\ref{6}) can be removed by a more general field redefinition  
\be
g_{\mu\nu}(\epsilon, x)=g_{\mu\nu}(x)+ g_{\mu\nu,1}(x)+g_{\mu\nu, 2}(x)+\dots ,
\lb{8-0}
\ee
where the first correction term is
\be
g_{\mu\nu, 1}=2a_1(z)\epsilon^2 f_{\mu\nu}(x)+2\epsilon^4 \sum_{k,l\geq 0}(\ln\epsilon)^{l+1}\mu_{k,l}(z)G^kh^{(k,l)}_{\mu\nu}(x)\, ,
\lb{8-2}
\ee
and the only constraints are imposed on the trace, $\Tr f=-R$ and  $\Tr h^{(k,l)}=-Z^{(k,l)}, \ k,l\geq 0$.  The solution, as before,  is  up to a trace free tensor.  
Conditions (\ref{7'}) guarantee that the correction term (\ref{8-2}) is small.
The redefinition (\ref{8-0}) with the first term (\ref{8-2}) removes the divergences already present in the action (\ref{6}). These divergences are canceled against the variation of first, $\sqrt{g}$,  term in the action.
There, however,  may appear  new UV divergences when we expand $\sqrt{g}$ up to second order in $g_{1}$ and take into account the variation of $\sqrt{g}R$ and $\sqrt{g}Z^{(k,l)}$ terms in the first order in $g_1$.  These divergences are milder than the original ones. They can be removed by adding a ``second order'' term $g_2$ in the redefinition (\ref{8-0}).
 The condition for the  cancellation of new divergences  implies only a constraint on the trace of $g_2$, 
\be
\Tr g_2=-\frac{1}{4}(\Tr g_1)^2+\frac{1}{2}\Tr g_1^2-2a_1\epsilon^2\Tr(Eg_1)-2\epsilon^4 \sum_{k,l\geq 0} (\ln\epsilon)^{l+1}\mu_{k,l}(\epsilon) G^k\Tr(\delta Z^{(k,l)}g_1)\, ,
\lb{8-3}
\ee
where $E_{\mu\nu}=R_{\mu\nu}-\frac{1}{2}g_{\mu\nu}R$ is the Einstein tensor and  $\sqrt{g}\delta_g Z^{(k,l)}=\delta (\sqrt{g}Z^{(k,l)}/\delta g^{\mu\nu}$ is the metric variation of the logarithmically divergent terms.
If, after the redefinition (\ref{8-0})-(\ref{8-3}), there still  produces a new UV divergent term in the action, it can be further removed by adding a ``third order'' term to (\ref{8-0}) and so on until
the action becomes finite (notice, that at each next step the degree of UV divergence decreases since the  divergence in the previous step gets multiplied by a small factor). Clearly, at each order there exists a simple algebraic  procedure to construct the appropriate
$g_{\mu\nu, p}$ term in (\ref{8-0}), some ambiguity in adding a trace free tensor is always present since only the trace of $g_{\mu\nu, p}$ is constrained. Each such term  is a local covariant function of the metric.  For any finite $\lambda$ and $\gamma_{k,l}$ in (\ref{7''}) only a finite number of steps is needed to make action finite.
Any UV divergences lower than that of the cosmological constant, thus, can be removed by the proposed mechanism. The UV divergence of the cosmological constant then needs to be renormalized.

\bigskip

Some remarks are in order:

\medskip

1. Let us summarize the key points of our proposal. First of all, we use any regularization which involves a dimensionful
regularization parameter $\epsilon$. Second, we assume that the cosmological constant is the most UV divergent term
in the effective action. This is obviously the case in one-loop. That it is still valid in higher loops 
is our assumption,  although a very natural one. Then  almost all UV divergences in the action can be hidden in a metric redefinition.
What remains is the UV divergence of the cosmological constant itself.

\medskip

2. Statement D can be reformulated in terms of the conformal rescaling of the metric,
\be
g_{\mu\nu}(x,\epsilon)=\sigma(x,\epsilon)g_{\mu\nu}(x)\, ,
\lb{90}
\ee
where $\sigma(x,\epsilon)$ is uniquely determined by the condition of cancellation of the divergences,
\be
\sigma(x,\epsilon)=1+\sigma_1+\sigma_2+\dots \, , \ \ \sigma_1=\frac{1}{4}\Tr g_1\, , \ \ \sigma_2=\frac{1}{4}\Tr g_2\, .
\lb{100}
\ee
The ambiguity present in (\ref{8-0}) is completely fixed in (\ref{90})-(\ref{100}). We stress that the field redefinition (\ref{90})-(\ref{100}) removes UV divergences
in the action for any, not necessarily conformally flat,  physical metric $g_{\mu\nu}(x)$. In an asymptotic region, where spacetime is well approximated by a maximally symmetric metric,
the conformal factor $\sigma(x,\epsilon)$ in (\ref{90}) becomes independent of the coordinates $x$ and the metric redefinition (\ref{90})
is a simple rescaling. This is advantage of having a local field redefinition.

\medskip

3. Removing  the UV divergences in front of the Einstein-Hilbert term in the action is not absolutely necessary. 
On the contrary, for reproducing a correct form of the Bekenstein-Hawking entropy one may need to keep the UV divergence of Newton's constant untouched, see \cite{Solodukhin:2015hma}.
In this case both Statements C and D give the desired solution to the problem provided one imposes $\Tr f=0$.  The only required condition on the couplings is the second condition in (\ref{7'}) saying that the higher curvature terms have a lower UV divergence than the cosmological constant. At the end, in this scenario, one would have to renormalize two physical parameters: cosmological constant and Newton's constant.

\medskip

4. The off-shell quantities are known to be gauge dependent. Therefore, it might be desirable to use the on-shell conditions for which the physical quantities such as $S$-matrix are gauge independent.   The proposed mechanism can be easily combined with the on-shell condition to be imposed on metric $g_{\mu\nu}(x)$. With this condition 
the UV divergent terms that vanish on-shell will not contribute to (\ref{8-2}) while the variation of the terms linear in the on-shell condition will contribute to (\ref{8-3}).
This goes similarly to the discussion, made for instance in \cite{G}, that  variation of terms, vanishing on-shell, does not necessarily vanish.

\medskip

5. It is interesting to note that the redefined metric $g_{\mu\nu}(x,\epsilon)$ (\ref{8-0})-(\ref{8-3}) resembles the Fefferman-Graham expansion for the
asymptotically AdS metric $g_{\mu\nu}(x,\rho)$, $\rho$ here is the radial coordinate and it plays the role of a small parameter in the expansion.
The metric $g_{\mu\nu}(x,\rho)$ satisfies Einstein equations in the space with coordinates $(x,\rho)$. The decomposition of this metric in $\rho$ is widely used in the AdS/CFT correspondence. 
In particular, the holographic UV divergences are obtained by decomposing the volume term $\sqrt{\det g(x,\rho)}$ in powers of $\rho$, see \cite{FG}. This is similar to our construction.
It would be interesting to see whether this is more than just a  similarity.

\medskip

6. Any redefinitions of the metric considered above have that nice property that they do not affect any classical term in the effective action since the relevant contributions disappear after one takes the limit $\epsilon\rightarrow 0$. Moreover, if gravity couples to a renormalizable theory the same applies: a variation of the action of this theory under any metric redefinition of the sort we discussed produces a small, negligible  in the limit of small $\epsilon$, contribution. Thus, the metric redefinition does not produce any new UV divergences.
This is so provided the UV divergence of cosmological constant is the leading one in the complete theory\footnote{In certain supersymmetric extensions of Einstein gravity $a_0(\epsilon)$ may vanish, see \cite{Christensen:1980ee}. The proposed mechanism is not applicable to those theories.}.  On the other hand, it is known \cite{Deser} that when the quantum gravity couples to 
quantum matter there may appear new UV divergent terms for the matter fields which were otherwise absent. Although we do not see any immediate obstacles why 
those new divergent terms can not be removed using same (or similar) mechanism this problem requires a more careful analysis.

\medskip

7. Our proposal, based on the field redefinition (\ref{8-0})-(\ref{8-3}), deals with the UV divergences in the action. It would be interesting to see whether this field redefinition 
can be used to more practical things such as computation of scattering amplitudes for the gravitons. 
This, possibly,  may require to impose extra constraints on the metric redefinition thus restricting the ambiguity in adding a trace free tensor that we already mentioned.
A natural  question is whether $S$-matrix can be defined consistently in the present approach. Although  we shall not attempt to answer  in full this important question  in the present note, we make the following observations. First of all, it is known (see \cite{Witten:2001kn}) that  the notion of $S$-matrix is not that easy to introduce in the presence of a non-vanishing cosmological constant.
It may be  therefore more convenient to  think in terms of correlation functions rather than scattering amplitudes. Moreover that even in flat spacetime (cosmological constant is zero) 
the elements of $S$-matrix can be expressed in terms of $n$-point correlation functions using the standard LSZ construction.  The end points in the correlation functions are supposed to be  taken to the asymptotic region. In the case of gravitons the correlation functions
are obtained by computing the  variations of the complete quantum effective action (with the gauge-fixing and ghosts terms included) with respect to metric (the 2-point function, for example, is obtained by inversing the  quadratic variation). The action expressed in terms of the physical metric $g_{\mu\nu}(x)$ has only one UV divergent term (cosmological constant) so that in a variation with respect to this metric all other UV divergences are already hidden in the field redefinition.
On the other hand, as we have already pointed this out in Remark 2, in asymptotic region the two metrics $g_{\mu\nu}(x,\epsilon)$ and $g_{\mu\nu}(x)$ are related by a simple, $\epsilon$-dependent, rescaling. These observations make us to think  that almost all UV divergences in a correlation function of gravitons  (with all end points lying in the asymptotic regions) 
will be  removed if it is expressed in terms of the physical metric $g_{\mu\nu}(x)$. The only UV divergence left is in the cosmological constant.
Additionally, the passage between variations  with respect to these two metrics produces an extra, regular, $\epsilon$-dependent factor. 
It would be of course interesting to check these expectations in a concrete calculation.

\medskip

8. It is natural to ask whether the proposed mechanism may be useful in other non-renormalizable theories such as a $\sigma$-model with a potential.
The  important condition for this proposal to work is the existence of a term in the action whose UV divergence is dominating.
Additionally, this term should be consistent with the symmetries of the theory. The existence of such a term and the concrete realization of the mechanism should be considered in each particular case.

\medskip

9. Clearly, the mechanism suggested in this note  can be generalized to any higher dimension $d>4$. The only property which is required is that the UV divergence of the cosmological constant  should be the highest
in the action.
Then the appropriate field redefinition removes all other divergences in the action. The only thing that remains is to renormalize the cosmological constant itself. 

\bigskip

In conclusion,  we have suggested that a (local)  metric redefinition  can be used to reduce the whole infinite set of UV divergences in the quantum  action for the Einstein gravity to a single UV divergence of  the cosmological constant.

\bigskip

%\section*{Acknowledgements} 
{\bf Acknowledgments:} I am grateful to D. Kazakov who, long  ago, introduced me to his work \cite{Kazakov:1987ej}. 
It is a  pleasure to thank A. Barvinsky, G. Gibbons and A. Tseytlin for useful remarks on the draft of this note.

\end{document}